\documentclass[prd,twocolumn]{revtex4}
\usepackage{dcolumn}
\usepackage{multirow}
\usepackage{graphicx}
\usepackage{amssymb}
\usepackage{bm}
\usepackage{hyperref}
\usepackage{epstopdf}
\usepackage{color}
\usepackage{mathrsfs}
\usepackage{amsmath,amssymb,amsthm}
\usepackage{rotating}
\usepackage{sverb, longtable}
\usepackage{subfigure}

\usepackage[graphicx]{realboxes}
\usepackage{adjustbox}

\begin{document}

\title{Constraining Cosmological Physics with DESI BAO Observations}

\author{Deng Wang}
\email{dengwang@ific.uv.es}
\affiliation{Instituto de F\'{i}sica Corpuscular (CSIC-Universitat de Val\`{e}ncia), E-46980 Paterna, Spain}

\begin{abstract}
The DESI year one observations can help probe new physics on cosmological scales. In light of the latest DESI BAO measurements, we constrain five popular cosmological scenarios including inflation, modified gravity, annihilating dark matter and interacting dark energy. Using a data combination of BICEP/Keck array, cosmic microwave background and DESI, we obtain the $1\sigma$ and $2\sigma$ constraints on the tensor-to-scalar ratio $r_{0.05}= 0.0176^{+0.0070}_{-0.0130}$ and $r_{0.05}=0.018^{+0.020}_{-0.017}$ indicating a beyond $2\sigma$ evidence of primordial gravitational waves. Using the combination of cosmic microwave background and DESI, we find a $2.4\sigma$ evidence for gravitational theories beyond the general relativity, shrink the dark matter annihilation cross-section by $12\%$ relative to cosmic microwave background, obtain a $1.3\sigma$ hint of the positive interaction between dark matter and dark energy implying that energy may be transferred from dark matter to dark energy in the dark sector of the universe, and give a clue of massive sterile neutrinos via the $2\sigma$ constraint on the effective number of relativistic degrees of freedom $N_{eff}=3.16^{+0.26}_{-0.11}$ and the effective mass $m^{eff}_{\nu, sterile}<0.52$ eV.
Future DESI observations could go a step further to explore the nature of inflation, dark matter, dark energy and neutrinos, and test the validity of general relativity on cosmological scales.

\end{abstract}
\maketitle

\section{Introduction}
Since the cosmic acceleration is discovered by Type Ia supernovae (SN) \cite{SupernovaSearchTeam:1998fmf,SupernovaCosmologyProject:1998vns} and confirmed by two independent probes cosmic microwave background (CMB) \cite{Planck:2018vyg,WMAP:2003elm,Planck:2013pxb} and baryon acoustic oscillations (BAO) \cite{SDSS:2005xqv,2dFGRS:2005yhx,Tully:2022rbj}, the standard 6-parameter cosmology, $\Lambda$-cold dark matter ($\Lambda$CDM) has achieved great success in depicting the physical phenomena across multiple scales at the background and perturbation levels from early times to late times. However, this model is imperfect and faces at least two problems, i.e., the so-called cosmological constant problem and the coincidence problem \cite{Weinberg:1988cp}. The former is why the value of cosmological constant from current observations is much smaller than that from the theoretical prediction, while the latter states why present-day densities of dark energy and dark matter are of the same order. $\Lambda$CDM also suffers from the small-scale problems such as the ``Too-big-to-fail'' crisis, which indicates that the dark matter mass derived from stellar kinematics of stars within the half-light radius for the most massive observed satellites of the Milk Way (MW) galaxy is smaller than those of massive subhalos in the $\Lambda$CDM cosmological simulations of our MW galaxy \cite{Boylan-Kolchin:2011qkt,Boylan-Kolchin:2011lmk}. Meanwhile, $\Lambda$CDM faces two main cosmological tensions emerged from recent cosmological observations, i.e., the Hubble constant ($H_0$) and matter fluctuation amplitude ($\sigma_8$) tensions \cite{DiValentino:2020vhf,Abdalla:2022yfr,DiValentino:2020zio}. The $H_0$ tension is that the indirectly derived $H_0$ value from the Planck-2018 CMB observations under $\Lambda$CDM is over 4$\sigma$ level lower than the direct measurement of present-day cosmic expansion rate from the Hubble Space Telescope (HST) \cite{Planck:2018vyg,Riess:2021jrx}, while the $\sigma_8$ one is that today's matter fluctuation amplitude in linear regime measured by several low redshift probes including weak gravitational lensing \cite{Planck:2018lbu,Heymans:2020gsg,KiDS:2020suj,KiDS:2020ghu,DES:2021vln,DES:2022ccp}, redshift space distortions \cite{Macaulay:2013swa} and cluster counts \cite{Battye:2014qga} is lower than that indirectly derived by the Planck-2018 CMB data \cite{Planck:2018vyg} under $\Lambda$CDM. It is very logically reasonable to query the validity of $\Lambda$CDM in describing the background evolution and structure formation of the universe. So far, to solve these problems, there are a large number of alternative models proposed by many authors (see \cite{Abdalla:2022yfr,DiValentino:2020zio} for reviews). It is noteworthy that, except the theoretical developments, more importantly, we require new high-precision observations. Particularly, we need independent and powerful probes with better accuracies to give definite answers on some key puzzles. To achieve this goal, a very promising probe is BAO. 

BAO is a pattern of wrinkles in the density distribution of galaxies spread across the universe, caused by acoustic density waves in the primordial plasma of the early universe. In cosmology, BAO provides a standard ruler for a typical length scale, which is the maximum distance that the acoustic waves could travel in the primordial plasma before the plasma cooled to the point where it became neutral atoms at the recombination epoch. BAO can be measured by analyzing the spatial distribution of galaxies and consequently provides an independent way to measure the expansion rate of the universe and how that rate has changed throughout cosmic history. 

Most recently, the DESI collaboration presents new high-precision BAO measurements \cite{DESI:2024uvr,DESI:2024lzq} and new cosmological results \cite{DESI:2024mwx} from the measurements of BAO in galaxy, quasar and Lyman-$\alpha$ forest tracers from the first year of observations from the Dark Energy Spectroscopic Instrument (DESI). DESI BAO provides robust measurements of the transverse comoving distance and Hubble rate \cite{DESI:2024mwx}, or their combination, relative to the sound horizon, in seven redshift bins from over 6 million extragalactic objects in the redshift range $z\in[0.1,\,4.2]$. In light of the DESI BAO data, we aim at probing new physics as completely as possible on cosmic scales. Specifically, we constrain five mainstream cosmological scenarios with the latest DESI data including inflation, modified gravity (MG), dark matter, dark energy and sterile neutrino models. By adopting the data combination of BK18+CMB+DESI, we obtain the $1\sigma$ and $2\sigma$ constraints $r_{0.05}= 0.0176^{+0.0070}_{-0.0130}$ and $r_{0.05}=0.018^{+0.020}_{-0.017}$, which gives the $2\sigma$ lower bound for the first time. The combined dataset CMB+DESI shows a $2.4\sigma$ deviation from the general relativity (GR), compresses the parameter range of the dark matter annihilation efficiency, provides a $1.3\sigma$ hint of the positive interaction between dark matter and dark energy and gives a clear clue of massive sterile neutrinos.

This study is outlined in the following manner. In the next section, we review briefly the alternative cosmological models to be constrained with the DESI data. In Section III, we describe the data and methodology considered. In Section IV, we present the constraining results. Discussions and conclusions are exhibited in the final section.

\section{Cosmological models}
In this study, we consider five popular cosmological models of interests including inflation, MG, annihilating dark matter (ADM), interacting dark energy (IDE) and massive sterile neutrinos.   

\subsection{Inflation}
The standard paradigm predicts that the very early universe undergoes an exponential expansion between $10^{-35}$ s and $10^{-33}$ s since the big bang happens \cite{Martin:2013tda,Achucarro:2022qrl}. This exponential expansion is the so-called inflation, which determines the initial conditions of our universe. For the inflation case, we consider the simplest $\Lambda$CDM+$r$ model same as the Planck collaboration \cite{Planck:2018vyg,Planck:2018jri}, where $r$ is the tensor-to-scalar ratio and characterizes the tensor mode of the primordial universe. This tensor mode is also known as primordial gravitational waves, which source a distinctive curl-like pattern in the CMB polarization and add additional power to the large-scale temperature power spectrum. We assume the tensor-mode spectrum is close to scale invariant, with spectral index given by the inflation consistency relation to second order in slow-roll parameters. The pivot scale we use is $k_p=0.002$ Mpc$^{-1}$.

\subsection{Modified gravity}
As is well known, the standard cosmology lies in the framework of GR. However, one important question is whether GR is valid on cosmic scales or not \cite{Clifton:2011jh,Koyama:2015vza,CANTATA:2021ktz}. In this study, we consider a model-independent growth parameterization to depict the gravity sector of the universe on cosmic scales \cite{Planck:2018vyg,Pogosian:2010tj}. We take the perturbed Friedmann-Lema\^{i}tre-Robertson-Walker metric (FLRW) given by the line element 
\begin{equation}
{\rm d}s^2 = -(1+2 \Phi) {\rm d}t^2 + a^2 (1-2\Psi) {\rm d}x^2~, \label{eq:metric}
\end{equation}
where $\Phi$ and $\Psi$ are the gauge-invariant gravitational potentials, which
are very nearly equal at late times in $\Lambda$CDM, and $a$ is the scale factor.

Under the assumption of $\Lambda$CDM, we adopt two phenomenological functions $\mu(k,a)$ and $\eta(k,a)$ to characterize the modified Poisson equation and effective anisotropic stress as follows \cite{Planck:2018vyg,Pogosian:2010tj}
\begin{equation}
k^2 \Phi = -4\pi G a^2 \mu (k,a) \left[\rho\delta+3(\rho+P)\sigma\right], \label{eq:Poisson} 
\end{equation}
\begin{equation}
k^2[\Psi-\eta(k,a)\Phi] = 12\pi G a^2 \mu (k,a)(\rho+P)\sigma, \label{eq:aniso}
\end{equation}
where $\rho$, $\delta$ and $P$ are the energy density, overdensity and pressure of each species, $\sigma$ is the anisotropic stress from relativistic species (photons and neutrinos), $k$ is the comoving wavenumber, and $G$ is the gravitational constant. Using effective quantities like $\mu(k,a)$  and $\eta(k,a)$ has the advantage that they are able to model any deviations from $\Lambda$CDM at the perturbation level. For simplicity, we only consider the time dependence of these two functions, since the scale dependence needs a better modeling of physics at nonlinear scales. Specifically, we take $\mu(a)$  and $\eta(a)$ as functions of the late-time dark energy density fraction $\Omega_{de}(a)$ \cite{Planck:2018vyg,Pogosian:2010tj}:
\begin{equation}
\mu(a)=1+E_{11}\Omega_{de}(a), \label{eq:mu} 
\end{equation}
\begin{equation}
\eta(a)=1+E_{22}\Omega_{de}(a), \label{eq:eta} 
\end{equation}
where $E_{11}$ and $E_{22}$ are two parameters to characterize the deviation from $\Lambda$CDM. It is clear that $\mu_0=\mu(a=1$) and $\eta_0=\eta(a=1)$ and $(\mu_0, \eta_0)$ is equivalent to $(E_{11}, E_{22})$ today. Hence, we need to constrain the parameter pair $(\mu_0, \eta_0)$ with observations in this model-independent growth parameterization.

\subsection{Annihilating dark matter}
Current observations prefer CDM that is cold and pressureless, clusters, and participate only in gravity interaction. Nonetheless, there is a possibility that dark matter may have self-interaction, i.e., two dark matter particles may annihilate \cite{Hooper:2018kfv}. CMB anisotropies are sensitive to such an exotic energy injection in the inter-galactic medium (IGM) and consequently provide a constraint on dark matter annihilation efficiency during the evolution of the universe. High-precision BAO measurements can help break the degeneracies between different cosmological parameters. Therefore, in light of the latest DESI data, we aim at implementing the constraints on the ADM model. 

Considering a self-annihilating dark matter particle $\chi$ and its antiparticle $\bar{\chi}$, the total rate of energy release per unit volume is shown as \cite{Chluba:2009uv}
\begin{equation}
\frac{{\rm d}E}{{\rm d}t}\bigg\rvert_{\chi\bar{\chi}}=2f_{\rm d}M_\chi c^2\langle\sigma v\rangle N_\chi N_{\bar{\chi}}, \label{eq:dEdt} 
\end{equation}
where $M_\chi\equiv M_{\bar{\chi}}$ denotes the mass of dark matter particle and its antiparticle, $c$ is the speed of light, $\langle\sigma v\rangle$ is the thermally averaged product of the cross-section and relative velocity of the annihilating dark matter particles, and $N_\chi \equiv N_{\bar{\chi}}=N_{\chi,0}(1+z)^3$ represents the number density of dark matter particles and their corresponding antiparticles with today's value $N_{\chi,0}\simeq 7\times10^{-9}{\rm cm}^{-3}\left(\frac{\Omega_\chi h^2}{0.13}\right)\left(\frac{M_\chi c^2}{100{\rm GeV}}\right)^{-1}$. $f_{\rm d}$ is the fraction of released energy that will be deposited into IGM. In general, one expects that $f_{\rm d}<1$ for all the annihilating channels. Here we assume $f_{\rm d}$ is a constant, since CMB anisotropies are very weakly sensitive to the redshift dependence of the transferred energy fraction.  

To carry out the numerical analysis easily, we rewrite Eq.(\ref{eq:dEdt}) as \cite{Chluba:2009uv}
\begin{equation}
\frac{{\rm d}E}{{\rm d}t}\bigg\rvert_{\chi\bar{\chi}}=\epsilon_0 f_{\rm d}N_{\rm H}(1+z)^3 {\rm eV\,s}^{-1}, \label{eq:dEdtnew} 
\end{equation}
where the dimensionless parameter $\epsilon_0$ reads as \cite{Chluba:2009uv}
\begin{widetext}
\begin{equation}
\epsilon_0\equiv1.5\times 10^{-24} \left(\frac{\Omega_\chi h^2}{0.13}\right)^2\left(\frac{M_\chi c^2}{100{\rm GeV}}\right)^{-1}\left[\frac{\langle\sigma v\rangle}{3\times 10^{-26} {\rm cm}^3 {\rm s}^{-1}}\right], \label{eq:epsilon0} 
\end{equation}
\end{widetext}
where we have used $N_{\rm H}\simeq1.9\times 10^{-7}{\rm cm}^{-3}(1+z)^3$ as the number density of hydrogen nuclei during the evolution of the universe.

\subsection{Interacting dark energy}
There is two main components in the universe, namely dark matter and dark energy. So far, there is no observations to rule out such a possibility that dark energy interacts phenomenologically with dark matter on cosmic scales. As a consequence, we shall consider the IDE scenario in this analysis. In this model, the covariant conservation of energy-momentum tensors of dark energy and dark matter reads as \cite{Wang:2016lxa,Yang:2018euj,Li:2023fdk,DiValentino:2017iww}
\begin{equation}
\nabla_\nu T^{\nu}_{\,\,\mu,\,de}=-\nabla_\nu T^{\nu}_{\,\,\mu,\,c}=Q_\mu,    \label{eq:conservation} 
\end{equation} 
where $Q_\mu$ is the energy-momentum transfer vector. 

The background dynamics of the IDE model is governed by the following equations
\begin{equation}
\rho'_{c}+3\mathcal{H}\rho_{c}=-aQ,    \label{eq:cdm} 
\end{equation} 
\begin{equation}
\rho'_{de}+3\mathcal{H}(1+\omega_{de})\rho_{de}=aQ,    \label{eq:de} 
\end{equation} 
where $\rho_x$ ($x=c, \, de$) denotes the energy density of each species, $\omega_{de}$ is the equation of state of dark energy, $\mathcal{H}$ is the conformal Hubble parameter, $Q$ is the energy transfer rate determined by $Q_\mu$, and the prime represents the derivative respect to the conformal time. 

At the linear perturbation level, we take the eigenfunctions of the Laplace operator $X$ and its covariant derivatives $X_i=-k^{-1}\nabla X_i$ and $X_{ij}=(k^{-2}\nabla_i\nabla_j+\gamma_{ij}/3)X$ with $\gamma_{ij}$ the spatial metric and $-k^2$ the eigenvalue of the Laplace operator, to express the FLRW metirc with scalar perturbations \cite{Kodama:1984ziu,Bardeen:1980kt} as 
\begin{equation}
g_{00}=-a^2(1+2AX), \quad g_{0i}=-a^2BX_i ,   \label{eq:pmetric1} 
\end{equation} 
\begin{equation}
g_{ij}=a^2(\gamma_{ij}+2W_LX\gamma_{ij}+2W_TX_{ij}),    \label{eq:pmetric2} 
\end{equation} 
where $A$, $B$, $W_L$, and $W_T$ depict the amplitudes of scalar metric perturbations.

Using Eqs.(\ref{eq:conservation}-\ref{eq:pmetric2}), the linearly perturbed equations of dark matter and dark energy can be expressed as
\begin{widetext}
\begin{equation}
\delta\rho'_x+3\mathcal{H}(\delta\rho_x+\delta p_x)+(\rho_x+p_x)(kv_x+3W'_L)=a(AQ_x+\delta Q_x),    \label{eq:delta} 
\end{equation} 
\begin{equation}
\left[\theta_x(\rho_x+p_x)\right]'+4\mathcal{H}(\rho_x+p_x)\theta_x-\delta p_x-A(\rho_x+p_x)+\frac{2}{3}\left(1-\frac{3K}{k^2}\right)p_x\sigma_x=a(\theta_xQ+\frac{f_x}{k}),    \label{eq:theta} 
\end{equation} 
\end{widetext}
where $\delta_x$, $p_x$ and $\theta_x\equiv (v_x-B)/k$ denote the perturbations of energy density, pressure and velocity for each species, respectively, $K$ is the spatial curvature, $\delta Q$ is the energy transfer perturbation, and $f_x$ is the momentum transfer rate for each component. It is worth noting that a positive $Q$ means that the energy and momentum transfers from dark matter to dark energy. In this study, we consider the IDE model $Q=\beta H_0\rho_c$, which is proportional to the energy density of dark matter. When the interaction strength $\beta=0$, the IDE reduces to $\Lambda$CDM.

\subsection{Sterile neutrinos}
Based on the fact that the solar and atmospheric neutrino oscillation experiments \cite{Super-Kamiokande:1998kpq,SNO:2003bmh} demonstrated that
the neutrinos have masses (see \cite{Lesgourgues:2006nd} for a recent review), massive neutrinos can potentially be one of the most appealing
solution to resolve $H_0$ and $\sigma_8$ discrepancies, because the free-streaming neutrinos can suppress the strength of matter clustering at late times. To be more specific, larger neutrino masses give a lower $\sigma_8$ and masses below 0.4
eV can provide an acceptable explanation to the directly local $H_0$ measurement (see also \cite{Planck:2018vyg} for details). Since the SK (Super-Kamiokande) and SNO (Sudbury Neutrino Observatory) experiments provide possible living space for extra sterile species \cite{Planck:2015fie}, sterile neutrinos can also be used to relieve the cosmological tensions \cite{Wyman:2013lza}. 

Generally, there are two classes of sterile neutrinos: (i) massless sterile neutrinos; (ii) massive sterile neutrinos. The former corresponds to the effective number of relativistic degrees of freedom $N_{eff}>3.046$, while the latter is that both a new kind of weakly interacting neutrinos with other species (such as photons) and three generation of actively massive neutrinos construct the cosmic neutrino background. The massive sterile neutrinos can affect the background evolution and structure formation of the universe. Thus, the DESI BAO data should be sensitive to massive sterile neutrinos in the neutrino sector. In this study, they are completely described by $N_{eff}$ and an effective mass parameter $m^{eff}_{\nu, sterile}$.

\begin{figure}
	\centering
	\includegraphics[scale=0.5]{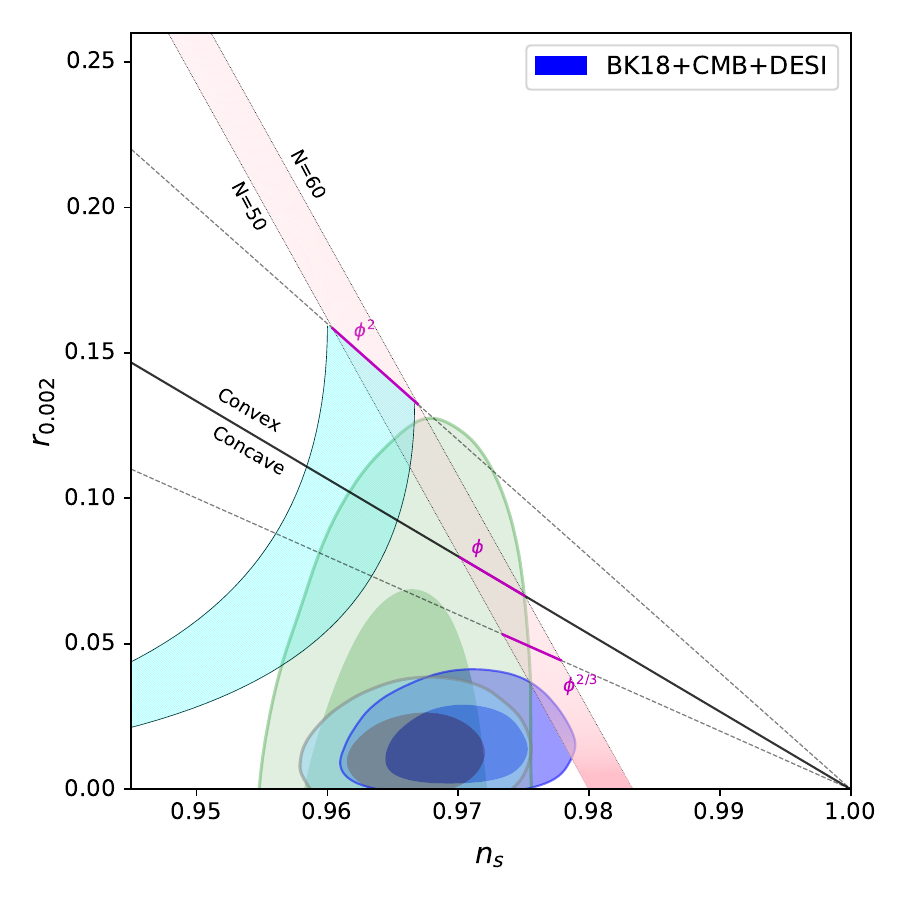}
	\caption{The two-dimensional marginalized posterior distributions of the parameter pair ($n_s$, $r_{0.002}$) from CMB (green), BK18+CMB+SDSS (grey) and BK18+CMB+DESI (blue) observations in the $\Lambda$+$r$ scenario, compared to the theoretical predictions of selected inflationary models. Here $\phi$ denotes the inflationary scalar field and $N\equiv{\rm ln}a$ is the e-folding number. }\label{fig:inflation}
\end{figure}

\begin{figure}
	\centering
	\includegraphics[scale=0.5]{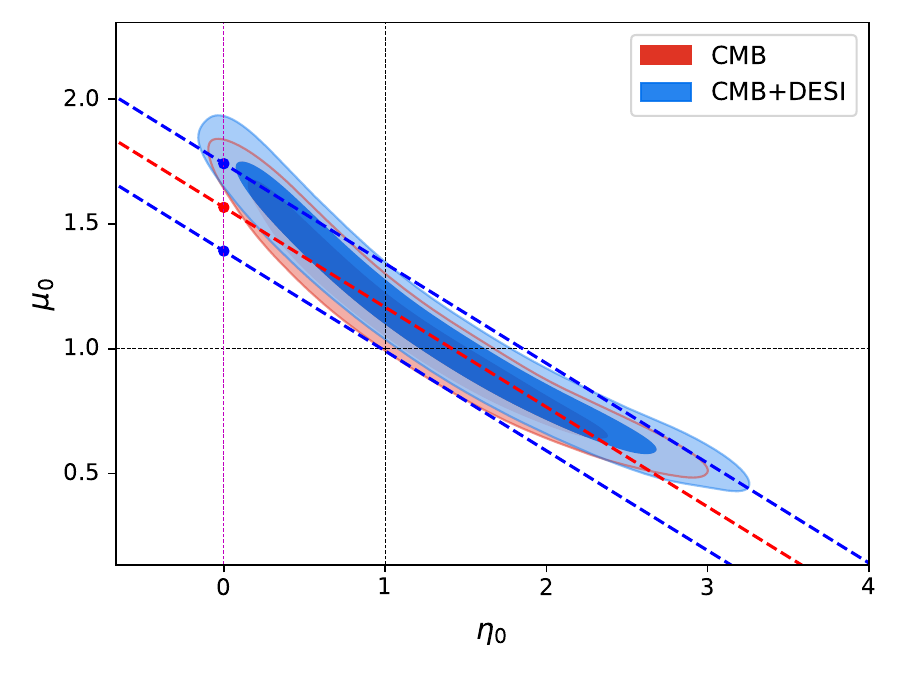}
	\caption{The two-dimensional marginalized posterior distributions of the parameter pair ($\mu_0$, $\eta_0$) from CMB (red) and CMB+DESI (blue) observations in the MG scenario. $S_0$ acts as a signal parameter to measure the deviation from the GR. The red and blue points denote the best fit value and $2\sigma$ limits of $S_0$, respectively. Similarly, the red and blue lines are the best fit line and $2\sigma$ boundaries when using the fitting formula $S_0=\mu_0+0.4\eta_0$, respectively. The magenta dashed line corresponds to $\eta_0=0$ and the cross point between black dashed lines represents the GR case.}\label{fig:MG}
\end{figure}

\begin{figure*}
	\centering
	\includegraphics[scale=0.35]{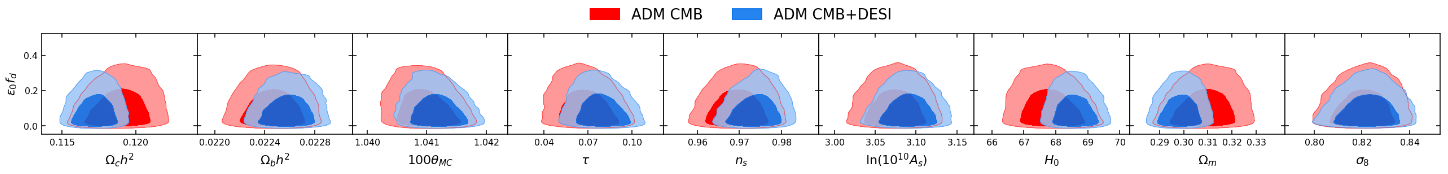}
	\caption{The two-dimensional marginalized posterior distributions of the cosmological parameters from CMB (red) and CMB+DESI (blue) observations in the ADM scenario.}\label{fig:ADM}
\end{figure*}

\begin{figure}
	\centering
	\includegraphics[scale=0.55]{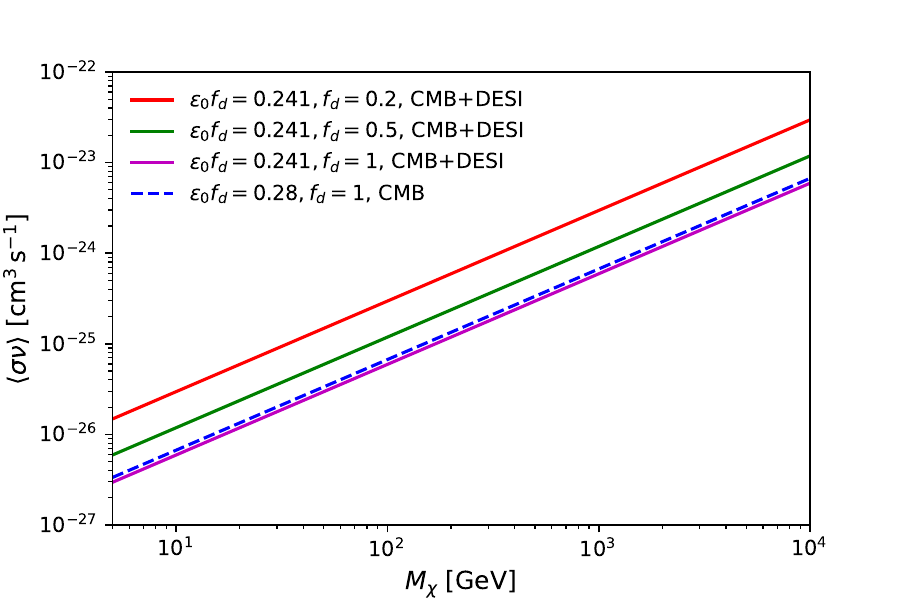}
	\caption{CMB+DESI constraints on the dark matter mass and annihilation cross-section. We assume the dark matter annihilation efficiency $\epsilon_0 f_{\rm d}=0.241$ and vary the transferred energy fraction $f_{\rm d}$. As a comparison, the dashed line denotes the Planck CMB bound assuming $\epsilon_0 f_{\rm d}=0.28$.}\label{fig:ADMcross}
\end{figure}

\begin{figure}
	\centering
	\includegraphics[scale=0.5]{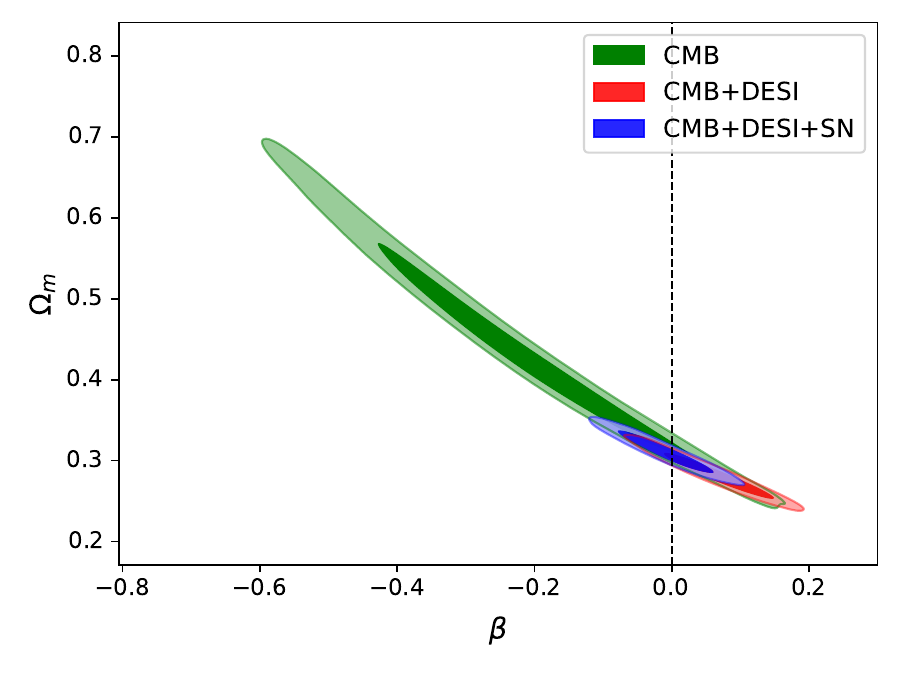}
	\caption{The two-dimensional marginalized posterior distributions of the parameter pair $(\beta, \Omega_m)$ from CMB (green), CMB+DESI (red) and CMB+DESI+SN (blue) observations in the IDE scenario. The vertical dashed line denote the $\Lambda$CDM case corresponding to $\beta=0$.}\label{fig:IDE}
\end{figure}

\begin{figure}
	\centering
	\includegraphics[scale=0.5]{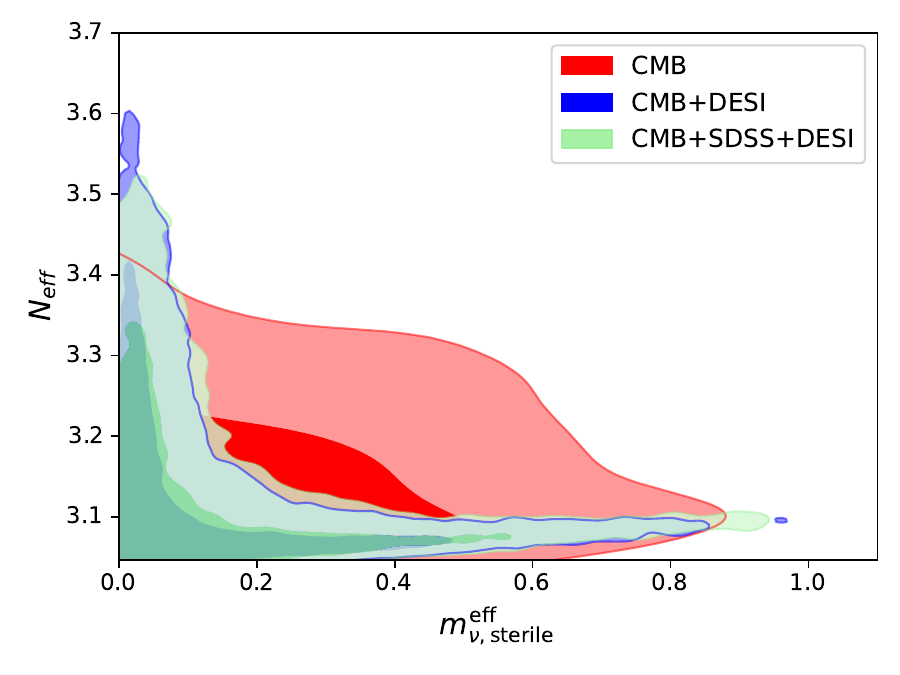}
	\caption{The two-dimensional marginalized posterior distributions of the parameter pair $(m^{eff}_{\nu, sterile},\, N_{eff})$ from CMB (red), CMB+DESI (blue) and CMB+SDSS+DESI (green) observations in the case of massive sterile neutrinos.}\label{fig:snu}
\end{figure}

\section{Data and methodology}
To study new physics beyond $\Lambda$CDM, we use the following observational datasets:

$\bullet$ CMB. Observations from the Planck satellite have very important meanings for cosmology and astrophysics. They have measured the matter components, the topology and the large scale structure of the universe. We adopt the Planck 2018 high-$\ell$ \texttt{plik} temperature (TT) likelihood at multipoles  $30\leqslant\ell\leqslant2508$, polarization (EE) and their cross-correlation (TE) data at $30\leqslant\ell\leqslant1996$, and the low-$\ell$ TT \texttt{Commander} and \texttt{SimAll} EE likelihoods at $2\leqslant\ell\leqslant29$ \cite{Planck:2019nip}. We also employ conservatively the Planck lensing likelihood \cite{Planck:2018lbu} from \texttt{SMICA} maps at $8\leqslant\ell \leqslant400$.

$\bullet$ BAO. BAO are very clean probes to explore the evolution of the universe, which are unaffected by uncertainties in the nonlinear evolution of the matter density field and by other systematic uncertainties which may affect other observations. Measuring the positions of these oscillations in the matter power spectrum at different redshifts can place strong constraints on the cosmic expansion history. Specifically, we use 12 DESI BAO measurements specified in Ref.\cite{DESI:2024mwx}, including the BGS sample in the redshift range $0.1 < z < 0.4$, LRG samples in $0.4 < z < 0.6$ and $0.6 < z < 0.8$, combined LRG and ELG sample in $0.8 < z < 1.1$, ELG sample in $1.1 < z < 1.6$, quasar sample in $0.8 < z < 2.1$ and the Lyman-$\alpha$ Forest Sample in $1.77 < z < 4.16$ \cite{DESI:2024uvr,DESI:2024lzq,DESI:2024mwx}. We also adopt the SDSS BAO measurements extracted from the 6dFGS \cite{Beutler:2011hx}, MGS \cite{Ross:2014qpa}, BOSS DR12 \cite{BOSS:2016wmc} and eBOSS DR16 \cite{eBOSS:2020yzd,eBOSS:2020mzp} samples.

$\bullet$ SN. Supernovae Ia luminosity distances are powerful distance indicators to probe the background evolution of the universe, especially, the equation of state of dark energy. We adopt SN data points from the Pantheon+ sample \cite{Scolnic:2021amr}, which consists of 1701 light curves of 1550 spectroscopically confirmed SNe Ia coming from 18 different surveys.

$\bullet$ BK18. To constrain the inflation physics, we use the BICEP2, Keck Array and BICEP3 CMB polarization experiments up to and including the 2018 observing season \cite{BICEP:2021xfz}. This datasets includes the additional Keck Array observations at 220 GHz and BICEP3 observations at 95 GHz relative to the previous 95/150/220 GHz data set. We refer to this dataset as ``BK18''.

In order to perform the Bayesian analysis, we employ the publicly available Boltzmann solver \texttt{CAMB} \cite{Lewis:1999bs} and take the Monte Carlo Markov Chain (MCMC) method to infer the posterior distributions of model parameters by using the public package \texttt{CosmoMC} \cite{Lewis:2002ah,Lewis:2013hha}. We analyze the MCMC chains via the online package \texttt{Getdist} \cite{Lewis:2019xzd}. The convergence diagnostic of the MCMC chains is the Gelman-Rubin quantity $R-1\lesssim 0.03$ \cite{Gelman:1992zz}.  We take the following uniform priors for model parameters: the baryon fraction $\Omega_bh^2 \in [0.005, 0.1]$, the cold dark matter fraction CDM fraction $\Omega_ch^2 \in [0.001, 0.99]$, the acoustic angular scale at the recombination epoch $100\theta_{MC} \in [0.5, 10]$, the amplitude of primordial power spectrum $\mathrm{ln}(10^{10}A_s) \in [2, 4]$, the scalar spectral index $n_s \in [0.8, 1.2]$, the optical depth $\tau \in [0.01, 0.8]$, $r \in [0, 3]$, the effective gravitational strength $\mu_0 \in [-3, 3]$, the gravitational slip $\eta_0 \in [-3, 3]$, the dark matter annihilation efficiency $10^{-23}\epsilon_0 f_{\rm d} \in [0, 1]$, $\beta \in [-1, 1]$, $N_{eff} \in [3.046, 10]$ and $m^{eff}_{\nu, sterile} \in [0, 3]$ eV.

\section{Results}
In light of the DESI year one data, our numerical results are presented in Figs.\ref{fig:inflation}-\ref{fig:snu}. For the inflation case, we obtain the constraint on the tensor-to-scalar ratio $r_{0.002}=0.0160^{+0.0063}_{-0.0120}$ at the $1\sigma$ (68\%) confidence level (CL) and $r_{0.002}=0.016^{+0.018}_{-0.016}$ at the $2\sigma$ (95\%) CL, using the combined datasets BK18+CMB+DESI. It is clear that this constraint is tighter than the $2\sigma$ upper bound $r_{0.002}<0.058$ from the Planck collaboration \cite{Planck:2018vyg}, although they use BK15 and previous SDSS BAO measurements. Very interestingly, BK18+CMB+DESI gives the $2\sigma$ lower bound 0. We derive the $1\sigma$ and $2\sigma$ constraints $r_{0.05}= 0.0176^{+0.0070}_{-0.0130}$ and $r_{0.05}=0.018^{+0.020}_{-0.017}$ at the pivot scale $k_p=0.05$ Mpc$^{-1}$. Although the $2\sigma$ upper limit 0.038 from BK18+CMB+DESI is slightly larger than 0.036 from the BICEP/Keck Collaboration using BK18+CMB+SDSS \cite{BICEP:2021xfz}, our constraint provides a clear $2\sigma$ lower limit 0.001. At the pivot scale $k_p=0.01$ Mpc$^{-1}$, we also derive the $1\sigma$ and $2\sigma$ constraints $r_{0.01}= 0.0168^{+0.0066}_{-0.0120}$ and $r_{0.01}=0.017^{+0.019}_{-0.017}$, which is weaker than the bound $r_{0.01}<0.028$ in Ref.\cite{Galloni:2022mok} from the data combination of BK18, Planck DR4 and LIGO-VIRGO gravitational wave observations. However, what is different is we find the $2\sigma$ lower bound 0 for $r_{0.01}$. Furthermore, BK18+CMB+SDSS+DESI gives the $1\sigma$ and $2\sigma$ constraints $r_{0.05}= 0.0170^{+0.0067}_{-0.0130}$ and $r_{0.05}=0.017^{+0.020}_{-0.017}$, which is just slightly tighter than the combined dataset BK18+CMB+DESI. It is noteworthy that BK18+CMB+DESI gives a larger spectra index $n_s = 0.9700\pm 0.0036$ than BK18+CMB+SDSS does \cite{BICEP:2021xfz}. This is because the DESI BAO data prefer a slightly higher cosmic expansion rate than the SDSS BAO data \cite{DESI:2024mwx}.  

In Fig.\ref{fig:inflation}, we show the theoretical predictions of several inflationary models. It is interesting that the $2\sigma$ contour from BK18+CMB+DESI has an overlap with the power-law inflationary scenarios which has a potential $V(\phi)\propto\phi^n$ (roughly $n<\frac{2}{3}$). This means such power-law inflationary models can be alive within $2\sigma$ CL in light of the DESI data.     

For the MG case, using the data combination of CMB+DESI, we obtain the $1\sigma$ constraints on the effective gravitational strength $\mu_0-1 = 0.10^{+0.31}_{-0.48}$ and the gravitational slip $\eta_0-1 = 0.33^{+0.68}_{-1.10}$. These results seems to demonstrate the validity of GR on cosmic scales, because they are well compatible with the point $(\mu_0, \eta_0)=(1, 1)$. Nonetheless, two-dimensional posterior distributions of the parameter pair $(\mu_0, \eta_0)$ shows clearly a deviation from GR using CMB or CMB+DESI data. Interestingly, DESI data pull the posterior distribution to be away from the GR case (see Fig.\ref{fig:MG}). In our previous work \cite{Wang:2023hyq}, we have proposed the signal parameter $S_0=\mu_0+0.4\eta_0$, which is a linear combination of $\mu_0$ and $\eta_0$, to depict the deviation from GR. When $S_0=1.4$, GR recovers. For the CMB-only case, we obtain $S_0=1.567\pm0.088$ after simple calculations, while $S_0=1.616\pm0.089$ for CMB+DESI. One can easily find that CMB and CMB+DESI are in $1.9\sigma$ \cite{Planck:2018vyg,Planck:2015bue} and $2.4\sigma$ tensions with GR, respectively. This indicates that DESI help improve the evidence of beyond GR.   

The DESI observations can also help probe the dark sector more precisely. In Fig.\ref{fig:ADM}, we present the constraining results of ADM using CMB and the combined dataset CMB+DESI. We obtain the $2\sigma$ upper bounds on the dark matter annihilation efficiency $\epsilon_0 f_{\rm d}<0.28$ for CMB and $\epsilon_0 f_{\rm d}<0.241$ for CMB+DESI. It is easy to derive that CMB+DESI compresses $14\%$ parameter range of $\epsilon_0 f_{\rm d}$. Since CMB+DESI (CMB) gives the best fitting value of dark matter fraction $\Omega_{c}h^2=0.11727$ (0.1189), these two datasets give close cross-sections in logarithmic parameter space (see Fig.\ref{fig:ADMcross}). However, when assuming the extreme case $f_{\rm d}=1$, we find that the cross-section from CMB+DESI shrinks by $12\%$ relative to that from CMB. Moreover, except DESI data leads to larger $H_0$ (smaller $\Omega_m$), it gives a larger $n_s$ and optical depth $\tau$. This is ascribed to the fact that DESI data prefers a larger cosmic expansion rate.   

The DESI data is very sensitive to the interaction between dark matter and dark energy in the cosmic dark sector. CMB gives the constraint on the IDE parameter $\beta=-0.22^{+0.17}_{-0.15}$ at $1\sigma$ CL, which is a loose constraint and suggests that energy is transferred from dark energy to dark matter on cosmic scales. However, CMB+DESI gives $\beta=0.065^{+0.056}_{-0.050}$, which indicates a positive energy transfer rate at $1.3\sigma$ CL. The addition of SN to CMB+DESI provides the $1\sigma$ constraint $\beta = -0.010\pm 0.046$ implying that there is no energy transferred from dark matter to dark energy. In Fig.\ref{fig:IDE}, one can easily find that the usage of SN data reduces the deviation from $\Lambda$CDM in the IDE model considered.

BAO measurements can place strong constraints on the massive sterile neutrinos. In Fig.\ref{fig:snu}, we obtain the $2\sigma$ constraints $N_{eff}=3.16^{+0.26}_{-0.11}$ and $(m^{eff}_{\nu, sterile}<0.52$ eV from CMB+DESI, which gives $\Delta N_{eff}\equiv N_{eff}-3.046>0$ at beyond the $2\sigma$ CL. However, the effective mass of such a kind of sterile neutrino is not determined and we just find an $2\sigma$ upper bound 0.52 eV. It is worth noticing that $(m^{eff}_{\nu, sterile}$ has a clear peak when it is less than 0.1 eV. Interestingly, to a large extent, the addition of DESI BAO data to CMB compresses the parameter space in the $(m^{eff}_{\nu, sterile},\, N_{eff})$ plane when compared to CMB alone \cite{Planck:2018vyg}. Furthermore, CMB+SDSS+DESI provides the $2\sigma$ constraints $N_{eff} = 3.14^{+0.23}_{-0.10}$ and $(m^{eff}_{\nu, sterile}<0.56$ eV. One can easily find that CMB+SDSS+DESI gives a slightly tighter constraint on $N_{eff}$ but a more loose constraint on $(m^{eff}_{\nu, sterile}$ than CMB+DESI does.

\section{Discussions and conclusions}
The newly released high-precision DESI BAO measurements can help investigate the background dynamics of the universe. In light of the DESI year one data, we attempt to probe new physics on cosmological scales. Specifically, we perform constraints on five popular cosmological models including inflation, modified gravity, annihilating dark matter, interacting dark energy and massive sterile neutrinos. For the inflation case, using the data combination of BK18+CMB+DESI, we obtain the $1\sigma$ and $2\sigma$ constraints $r_{0.05}= 0.0176^{+0.0070}_{-0.0130}$ and $r_{0.05}=0.018^{+0.020}_{-0.017}$ at the pivot scale $k_p=0.05$ Mpc$^{-1}$. To the best of our knowledge, this is the first time to give the $2\sigma$ lower bound in the literature. We obtain similar results by analyzing the one-dimensional posterior distributions of $r_{0.002}$ and $r_{0.01}$. The CMB+DESI data introduces a possibility that the power-law inflation is revived within $2\sigma$ CL, because it predicts a larger spectra index. 

Employing the signal parameter $S_0$, CMB+DESI gives a $2.4\sigma$ deviation from the GR, which is higher than $1.9\sigma$ given by CMB. Interestingly, this reveals that the DESI data prefers beyond GR physics in the gravity sector on cosmological scales. The addition of DESI to CMB leads to the fact that the dark matter annihilation cross-section shrinks by $12\%$ relative to that from CMB. It is worth noting that this improvement actually stands for all the $f_{\rm d}$ values. Therefore, DESI can help constrain the dark matter annihilation. Furthermore, constraining the IDE model with CMB+DESI observations, we obtain the constraints on the typical interaction parameter $\beta=0.065^{+0.056}_{-0.050}$, which is a $1.3\sigma$ hint and indicates that energy may be transferred from dark matter to dark energy on cosmic scales. However, the addition of SN to CMB+DESI causes the disappearance of this possible signal. Interestingly, CMB prefers a negative interaction between dark matter and dark energy. This consequence is very similar to the case of cosmic curvature. CMB alone gives a closed universe \cite{Planck:2018vyg} but CMB+DESI prefers an open universe \cite{DESI:2024mwx}. In addition, although CMB+DESI gives an evidence of nonzero effective number of massive sterile neutrinos beyond the $2\sigma$ CL $(N_{eff}=3.16^{+0.26}_{-0.11})$, we can not claim the evidence of massive sterile neutrinos due to the fact that its effective mass is still not determined at the $2\sigma$ CL $(m^{eff}_{\nu, sterile}<0.52)$ eV. 

The DESI survey will release more high-precision and more complete data in the following several years. To a large extent, these forthcoming observations would help explore the nature of inflation, dark matter and dark energy, and test the correctness of GR on cosmological scales.

\section{Acknowledgements}
DW warmly thanks Olga Mena and Eleonora Di Valentino for helpful communications and discussions.
DW is supported by the CDEIGENT Project of Consejo Superior de Investigaciones Científicas (CSIC).

\end{document}